\documentclass{optica-article}

\journal{opticajournal} 

\articletype{Research Article}

\usepackage{lineno}
\usepackage{graphicx}
\usepackage{subcaption}
\microtypesetup{expansion=false} 

\usepackage{hyperref}
\usepackage{cleveref} 
\usepackage{float}

\begin{document}

\title{Contact-free charge carrier density determination of doped sub-\textmu m thin cubic GaN layers with terahertz time-domain spectroscopy}

\author{Joshua Hennig\authormark{1,2,*}, Mario F. Zscherp\authormark{3}, Silas A. Jentsch\authormark{3}, Daniel Anders\authormark{3}, Jörg Schörmann\authormark{3}, Till Fuchs\authormark{3}, Anja Henss\authormark{3}, Sangam Chatterjee\authormark{3}, Christian Röder\authormark{4}, Franziska C. Beyer\authormark{4}, Christina Jörg\authormark{2}, Joachim Jonuscheit\authormark{1}, Georg von Freymann\authormark{1,5}, and Daniel Molter\authormark{1}}

\address{\authormark{1}Fraunhofer Institute for Industrial Mathematics ITWM, Department Materials Characterization and Testing, 67663\,Kaiserslautern, Germany\\
\authormark{2}Department of Physics and Research Center OPTIMAS, RPTU Kaiserslautern-Landau, 67663\,Kaiserslautern, Germany\\
\authormark{3}Institute of Experimental Physics I and Center for Materials Research, Justus Liebig University Giessen, 35392\,Giessen, Germany\\
\authormark{4}Fraunhofer Institute for Integrated Systems and Device Technology IISB, Department Energy Materials and Test Devices, 91058\,Erlangen, Germany\\
\authormark{5}Institute of Photonic Systems and Technologies, Leibniz University Hannover, 30060\,Hannover, Germany}

\email{\authormark{*}joshua.hennig@itwm.fraunhofer.de}


\begin{abstract*} 
As cubic GaN is a promising material for optoelectronic applications, its layers need to be precisely doped to realize the desired properties. While contact-based methods are established to measure the realized charge carrier density, contact-free techniques are more advantageous for this purpose. We show terahertz time-domain spectroscopy measurements of cubic GaN layers with thicknesses between 0.6~\textmu m and 2~\textmu m with doping densities in the range from $10^{18}$~cm$^{-3}$ to $10^{20}$~cm$^{-3}$. From these measurements, we determine the charge carrier densities and verify the results with time-of-flight secondary-ion mass spectrometry measurements. The results are in accordance with simulations of the doping ranges which the technique is expected to be sensitive in. In preparation for the terahertz time-domain spectroscopy measurements, we validated the layer thicknesses by Fourier-transform infrared spectroscopy measurements. With this, a wide range of doped cubic GaN samples can be addressed by optical methods. In the future, these techniques could allow for a fast and non-destructive characterization of GaN layers, right after or even during the growth process.

\end{abstract*}

\section{Introduction}

Wide-bandgap (WBG) semiconductors are becoming increasingly important for many applications, e.g. in power electronics applications ranging from electric vehicles over photovoltaic applications to energy storage and consumer electronics \cite{zhang2025wide, viewegh2026wide, kumar2022wide, chaudhary2023technology}. Gallium nitride (GaN) is one of the most widely used representatives of WBG semiconductors besides silicon carbide \cite{singh2022recent} and WBG oxide semiconductors \cite{shi2021wide}. In addition to the applications of GaN in power electronics \cite{musumeci2023gallium}, this material system has also become a cornerstone in optoelectronics, from first GaN-based blue light-emitting diodes (LED) \cite{pankove1972gan, ohki1981fabrication, nakamura1993p} to the sophisticated devices established nowadays, such as highly efficient LEDs of multiple colors and with high power densities \cite{nakamura2013history,li2024recent}.

Most of these well-engineered applications use the preferred and thermodynamically most stable hexagonal (i.e., wurtzite) phase of GaN (h-GaN) as it is the ground state of the bulk material, whereas cubic (i.e., zinc-blende) GaN (c-GaN) is in a metastable phase \cite{lazarov2005selected}. Therefore, the hexagonal phase has become widely used and commercialized over decades \cite{isamu1997crystal, scholz2012semipolar}, while the cubic phase has gained attraction more slowly \cite{moustakas1993growth, as1996p}. Nonetheless, it is a promising material with certain technological advantages over its hexagonal polytype. A possible advantageous application lies in the context of \textmu -LEDs. While the state-of-the-art material for red LEDs is AlGaInP \cite{gessmann2004high}, the efficiency suffers from miniaturization \cite{fan2023analysis}. Alloys based on c-GaN, such as In$_{x}$Ga$_{1-x}$N, can, in principle, cover large parts of the electromagnetic spectrum extending the visible range towards both the infrared and the ultraviolet regions. For h-GaN, however, internal fields cause a quantum-confined Stark effect, reducing the efficiency towards longer wavelengths, opening up an opportunity for cubic nitrides to be the material system of choice for red \textmu -LEDs \cite{jentsch2025metal}. Going a step further towards multi-color LEDs, it is desirable to stay within one material system, being another possible application for c-GaN \cite{lu2023red}.

The process of doping cubic GaN layers with Ge is a key step along the way to realize these devices. Yet, it is still subject to ongoing research and optimized progressively \cite{zscherp2022aln, jentsch2025metal}. The electrical properties, especially charge carrier density and mobility, are often determined by Hall measurements in Van der Pauw geometry or capacitance-voltage measurements. Both of these techniques require physical contacts, e.g., gold wires, to be soldered \cite{as2003growth} or deposited onto the sample \cite{hernandez2020study, vilchis2015electrical, as2013capacitance}.
So, contactless methods would be beneficial to allow a sample to be characterized without limiting its further use. Time-of-flight secondary-ion mass spectrometry (ToF-SIMS) is a method, which can be used to determine the overall dopant concentration, which can be correlated with the charge carrier density directly \cite{jentsch2025metal, deppe2019germanium}. An optical method to measure this latter quantity directly is spectroscopic ellipsometry, e.g. in the infrared or the ultraviolet \cite{baron2019influence}. While it has been shown that this is a possibility, the technique requires extensive modeling, fitting of the results and measurements under multiple angles, which increases measurement times. Instead, a promising contactless and non-destructive technique is terahertz time-domain spectroscopy (TDS), which is becoming more and more deployed to applications in the context of semiconductors \cite{true2021review} after having become established as a standard technique for the characterization of insulators, such as plastic or paint \cite{ellrich2020terahertz}. 

\section{Methods and materials}

\subsection{Terahertz time-domain spectroscopy}

In this work, we use a terahertz TDS setup in reflection geometry to determine the charge carrier density of cubic GaN samples. For TDS, a transmission setup is the most basic realization \cite{koch2023terahertz} and has originally been used for semiconductor characterization as well \cite{van1990carrier}. In transmission geometry, it allows to measure highly doped semiconductors only as long as they are thin-film samples on an insulating substrate \cite{ulatowski2020terahertz}. But in cases of doped samples, that are still fairly thin, but no longer considered a thin film anymore, or in the case of highly-doped substrates, a reflection geometry is preferred. This has been realized early on by Jeon \textit{et al.} \cite{jeon1998characterization} and has become competitive for certain applications, for example in the characterization of graphene \cite{mackenzie2018quality}. A challenge of the reflection geometry is the correct detection of the phase, which can easily be erroneous due to slight misalignments or position errors of the necessary reference measurement \cite{koch2023terahertz}. Attempts of reference-free characterization methods have been proposed in both transmission \cite{whelan2020reference} and reflection configurations \cite{wenzel2026self}. Another way to address this problem is to make use only of the amplitude \cite{zografopoulos2024terahertz}. An amplitude-based evaluation has already been successfully applied for the electrical characterization of silicon \cite{hennig2024wide} and silicon carbide \cite{hennig2025simultaneous}.

Hexagonal GaN samples have been measured with terahertz TDS in a transmission setup before \cite{nagashima2005measurement, balci2012characteristics, zhang2003terahertz, guo2009terahertz}. In the study by Nagashima \textit{et al.} \cite{nagashima2005measurement} for example, samples of thicknesses above 1~\textmu m were investigated and the authors faced challenges to determine the charge carrier density separately from mobility and a combined resistivity, especially for highly doped samples \cite{nagashima2005measurement}. Another paper regards a hexagonal GaN sample with a thickness of 4.8~\textmu m measuring the frequency-dependent complex conductivity at different temperatures \cite{balci2012characteristics}. Further, a study of even thicker, free-standing h-GaN also focused on the complex conductivity and dielectric function in the terahertz frequency range \cite{zhang2003terahertz}. Another similar study of approx. 2~\textmu m-thick h-GaN applying the Drude model to determine electrical properties exists \cite{guo2009terahertz}. A setup similar to TDS has been used to study InGaN/GaN quantum wells as sources of terahertz radiation and to determine the mobility of a 2D electron gas at low temperatures \cite{burmistrov2023terahertz}. 

Without claiming to represent an exhaustive review, these publications exemplarily show that there has been research concerning the characterization of GaN samples using TDS before, mostly applying transmission setups and primarily focusing on the hexagonal polytype. Yet, cubic GaN samples have mostly been neglected thus far, and especially GaN with thicknesses below 1~\textmu m -- which is where the desirable thicknesses start, e.g. for highly-doped substrates for growth of further layers with applications in optoelectronics in mind -- have not been characterized with TDS thus far. This is not surprising, given that for terahertz wavelengths of multiple hundreds of microns, this poses a challenging task. Further, mostly quantities such as conductivity or its inverse value, resistivity, which holds the information of charge carrier density and mobility combined, have been subject to these studies. However, it is highly desirable from a material-growing point of view to determine the charge carrier density non-destructively and separately from other quantities as it is the quantity directly affected by the amount of doping. 

\begin{figure}[htb!]
\centering
    \includegraphics[width=0.8\textwidth]
    {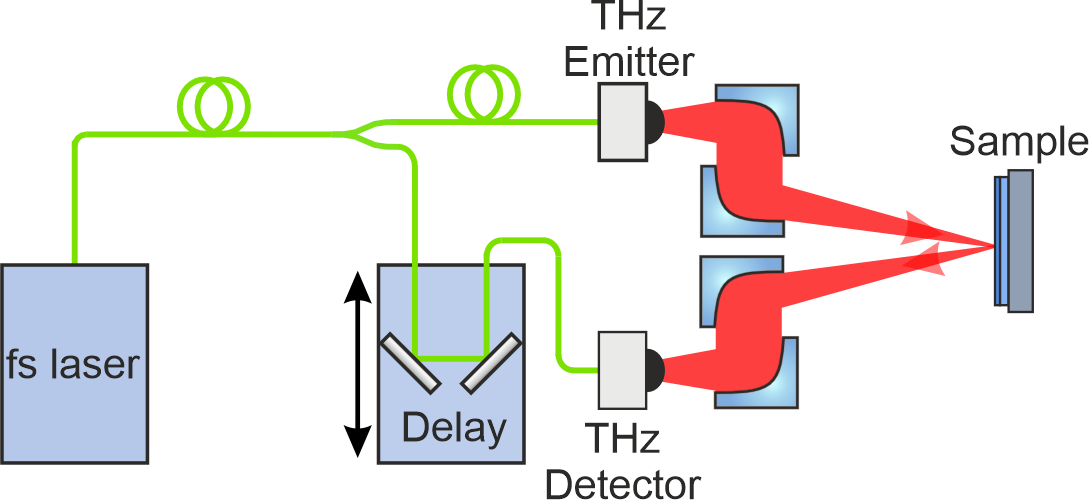}
    \caption{Schematic of a fiber-coupled terahertz TDS setup in reflection geometry. The beam of a fs laser is split into two arms. One of them is guided to the terahertz emitter and the other one via a delay unit to a terahertz detector. The terahertz beam is focused onto the sample by parabolic mirrors.}
    \label{fig:TDS}
\end{figure}


Fig.~\ref{fig:TDS} shows a schematic of a TDS setup in reflection geometry, which is applied in the following. A fs-laser pulse is split into two arms, guided to the terahertz emitter and detector, respectively. A ps-short terahertz pulse is emitted, whose propagation path is sketched in red, collimated and focused onto the sample by parabolic mirrors. After the interaction with the sample, it is guided towards the detector. Shifting the delay of the fs-pulse activating the detector allows to scan the complete terahertz waveform \cite{koch2023terahertz}. State-of-the art terahertz TDS systems applying electronically controlled optical sampling (ECOPS) reach measurement rates in the kilohertz range \cite{dietz2014all}.

Previous simulations predict that even for thin sample layers on the scale of 1~\textmu m and high charge carrier densities above 10$^{18}$~cm$^{-3}$  a characterization in a reflection geometry should be possible. This has been shown for various semiconductor materials, including hexagonal GaN \cite{hennig2026accessibility}. The applied reflection setup allows for the use of highly-doped substrates, which are widely used in the semiconductor industry making reflection TDS an even more desirable technique for applications in that field. However, it requires a charge carrier mobility model suitable for the material as well as precise knowledge of the individual layers' thicknesses, especially in the case of such thin layers. Since both the mobility and the sample's layer thicknesses are very important for a successful evaluation of the TDS measurements \cite{hennig2026accessibility}, the modeling of the mobility of the c-GaN samples will be discussed in more detail and the layer thicknesses are verified beforehand.

\subsection{Cubic Gallium nitride samples and characterization} 

\begin{figure}[htb!]
\centering
    \includegraphics[width=0.85\textwidth]
    {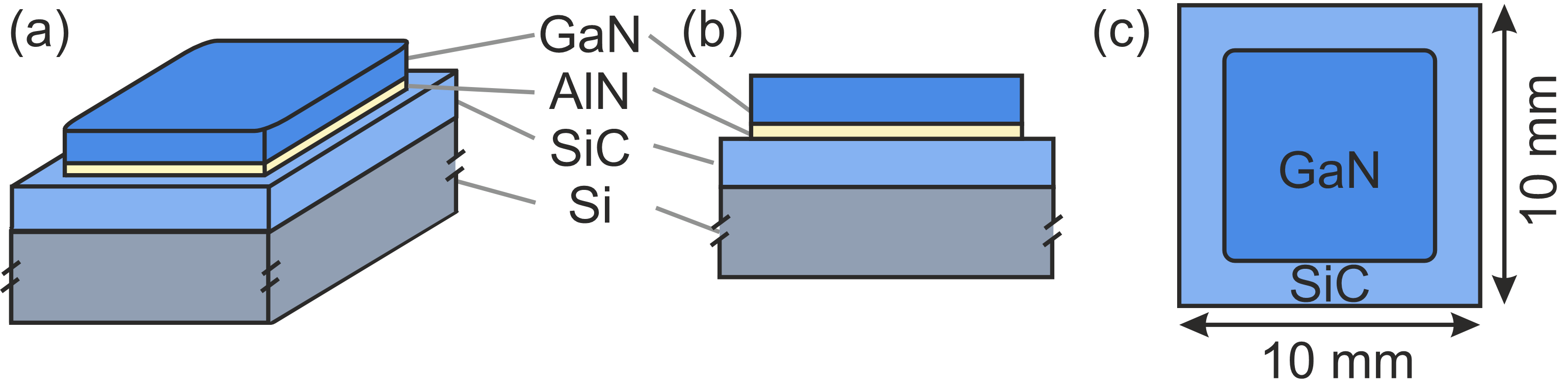}
    \caption{Sketches of the four-layer sample from different views: (a) 3D sketch, (b) side view, and (c) top view of a sample consisting of the layers c-GaN ($\approx$\,0.6\,\textmu m), AlN ($\approx$\,6\,nm), 3C-SiC ($\approx$\,10\,\textmu m), and Si ($\approx$\,525\,\textmu m) from top to bottom. Sketches are not to scale.}
    \label{fig:GaN_sample_sketch}
\end{figure}

The samples are composed of four layers as sketched in Fig. \ref{fig:GaN_sample_sketch}. Cubic GaN:Ge layers were grown by plasma-assisted molecular beam epitaxy on n-type 3C-SiC/Si pseudo substrates using a thin AlN buffer layer \cite{zscherp2022aln}. The pseudo substrate consists of CVD grown, n-type doped 3C-SiC with a thickness of 10.3~\textmu m and a doping concentration of a few $10^{17}$ cm$^{-3}$ deposited on 525~\textmu m thick lightly doped Si (001), according to the manufacturer's specification.

\begin{table}[b]
\caption{Reference sample information of the investigated c-GaN samples. Thickness $d_{\text{GaN}}$ information stems from the scanning electron microscopy characterization and Ge concentration $N_{\text{Ge}}$ in the GaN layer from ToF-SIMS measurements. 
Because the thickness values for S1 and S6 are not measured directly, but interpolated, a higher error is assumed in this case. Because the charge carrier density of the uid sample S1 does not stem from Ge-doping, but from possible background impurities, it could not be retrieved from ToF-SIMS measurements.}  
  \label{tab:sample_information}
  \centering
\begin{tabular}{ccc}
\hline
Number & $d_{\text{GaN}}$ / \textmu m & $N_{\text{Ge}}$ / cm$^{-3}$ \\ 
\hline
S1 & 2.27 $\pm$ 0.10 & uid \\
S2 & 0.57 $\pm$ 0.02 & 1$\cdot$10$^{18}$ $\pm$ 1$\cdot$10$^{18}$\\
S3 & 0.59 $\pm$ 0.02 & 4$\cdot$10$^{18}$ $\pm$ 2$\cdot$10$^{18}$\\
S4 & 0.60 $\pm$ 0.02 & 1$\cdot$10$^{19}$ $\pm$ 4$\cdot$10$^{18}$\\
S5 & 0.56 $\pm$ 0.02 & 4$\cdot$10$^{19}$ $\pm$ 2$\cdot$10$^{19}$\\
S6 & 1.00 $\pm$ 0.10 & 8$\cdot$10$^{19}$ $\pm$ 2$\cdot$10$^{19}$\\
\hline
\end{tabular}
\end{table}

Tab.~\ref{tab:sample_information} lists the samples investigated in this work in order of increasing charge carrier density. The reference values for the thicknesses $d_{\text{GaN}}$ are based on the characterization of the sample using cross-sectional scanning electron microscopy images. The Ge concentration $N_{\text{Ge}}$ is assessed based on ToF-SIMS measurements using an M6plus instrument (IONTOF GmbH, Münster, Germany) equipped with a 30 kV Bi liquid metal ion gun (LMIG), using Bi$^{+}$ species for analysis and a dual source column (DSC) using 500~eV or 1~keV Cs$^{+}$ for depth profiling. 
The analyzer was operated in spectrometry mode, yielding a mass resolution ($m/\Delta m$) of around 10,000 at $m/z$ = 19.00\,u (F$^{-}$). Charge compensation was maintained using a low-energy electron flood gun. An analysis area of $75 \times 75$~\textmu m$^{2}$ was rastered at a $128 \times 128$ pixel resolution with an $80$~\textmu s cycle time with intermittent $300 \times 300$~\textmu m$^{2}$ sputtering. Spectra were calibrated using Ga$^{-}$, GaN$^{-}$, and Ga$_{2}^{-}$ fragments. Data evaluation was conducted with Surface Lab 7.6 software package (IONTOF GmbH, Münster, Germany).

The concentration of Ge in the GaN:Ge layer is determined based on the intensity ratio of GeN$^-$ and GaN$^-$ fragments. For calibration, reference GaN:Ge samples with known concentration of Ge prepared by ion implantation are utilized \cite{deppe2019germanium}. The unintentionally doped (uid) c-GaN layer has an estimated charge carrier density roughly in the order of magnitude 1$\cdot$10$^{17}$ cm$^{-3}$. For the intentionally Ge-doped samples, the range of Ge concentrations realized is assessed more precisely by ToF-SIMS.

As nearly all Ge atoms in c-GaN are assumed to be electrically active as a donor and contribute one charge carrier each, the determined Ge concentration $N_{\text{Ge}}$ is suitable to be used as an estimate for the density of free charge carriers $n_{\text{GaN}}$ in the GaN layers \cite{deppe2019germanium}. 

\subsection{Drude and charge carrier mobility modeling}

For doped semiconductors, it has been proven to be sufficient to use the Drude model to model electrical conduction \cite{hennig2024wide,hennig2025simultaneous,hennig2026accessibility}. With this approach, the dielectric function $\tilde{\varepsilon}$ can be expressed as a function of charge carrier density $n$ and frequency $\omega$ by

\begin{equation}
    \tilde{\varepsilon}(n, \omega) = \varepsilon_{\text{S}}-\frac{ne^{2}}{\varepsilon_{0}m_{\text{eff}}m_{\text{e}}\omega \big  (\omega+\text{i}\frac{e}{m_{\text{eff}}m_{\text{e}}\mu(N)} \big )},
    \label{eq:dielectric_function_Drude_N_dependend}
\end{equation}

with the static dielectric constant $\varepsilon_{\text{S}}$, the elementary charge $e$, the vacuum permittivity $\varepsilon_{0}$, the effective mass $m_{\text{eff}}$, the electron mass $m_{\text{e}}$, the imaginary unit $\text{i}$, and the charge carrier mobility $\mu$. This latter quantity is again affected by the charge carrier density. For many materials, such as silicon \cite{hennig2024wide} or 4H-SiC \cite{hennig2025simultaneous}, this dependency can be modeled quite well by an empirical model, the Caughey-Thomas model \cite{caughey1967carrier}. This model can also be applied to 3C-SiC \cite{roschke2001electron}. 

However, because c-GaN is neither as widely used and established nor grown with low numbers of defects and dislocations as consistently as other semiconductor materials, there is no such sufficiently reliable empirical model available. In literature, one can find theoretical calculations for the lattice mobility \cite{lee2024high} and many publications reporting on both charge carrier density and mobility values for cubic GaN samples doped with Si \cite{kim1994growth, brandt1997growth, as1999heteroepitaxy, martinez2001p}. For p-type Mg-doped c-GaN, a collection of publications stating these two quantities is given in \cite{hernandez2020study} and further values can be found in \cite{vilchis2015electrical}. From all of these values combined, it is not easily possible to fit a Caughey-Thomas model, accounting for the various doping types and materials, the varying sample thicknesses, and mostly unreported dislocation and impurity densities which likely differ. Therefore, for the Ge-doped c-GaN samples investigated in this work, we model the mobility theoretically. 


According to \cite{schikora2002molecular, as2003growth}, the mobility $\mu$ of cubic GaN can be modeled analogously to that of hexagonal GaN by applying Matthiesen's rule

\begin{equation}
    \frac{1}{\mu} = \frac{1}{\mu_{\text{dsl}}} + \frac{1}{\mu_{\text{ii}}} + \frac{1}{\mu_{\text{latt}}}
    \label{eq:mobility_Matthiesens_rule}
\end{equation}

with separate contributions due to the influence of scattering at charged discolations $\mu_{\text{dsl}}$, and ionized impurities $\mu_{\text{ii}}$ as well as the lattice contribution $\mu_{\text{latt}}$ \cite{weimann1998scattering}.

Coulomb scattering at threading dislocations is one of the two main scattering mechanisms according to Weimann \textit{et al.} \cite{weimann1998scattering} that decrease the transverse mobility in GaN. Its contribution can be calculated via 

\begin{equation}
    \mu_{\text{dsl}} = \frac{\hbar^3 \varepsilon_{\text{S}}^2 \varepsilon_0^2 c^2}{e^3 f^2 m_{\text{eff}}^2 m_{\text{e}}^2 N_{\text{dsl}} \lambda_{\text{D}}^4} (1 + (2 k_{\perp} \lambda_{\text{D}})^2)^{3/2}
    \label{eq:mobility_discolations}
\end{equation}

with the reduced Planck constant $\hbar$, the lattice constant $c$, the fraction of filled traps $f$, the dislocation density $N_{\text{dsl}}$, the Debye length $\lambda_{\text{D}}$, and the transverse component of the electron wave vector $k_{\perp}$. The lattice constant for c-GaN is assumed to be $c = 0.451$\,nm \cite{okumura1997growth}. The calculation of $\mu_{\text{dsl}}$ is carried out according to \cite{weimann1998scattering}, replacing the hexagonal values by the cubic ones when appropriate.

To model the impact of ionized impurity scattering, Weimann applies the Conwell-Weisskopf formula \cite{conwell1950theory}. However, this model is further improved by the Brooks-Herring formula by introducing an exponential screening parameter \cite{ridley1977reconciliation}. Therefore, we use the Brooks-Herring formula according to \cite{li1977dopant, eggert2007beweglichkeit} describing the mobility contribution of ionized impurities via

\begin{equation}
    \mu_{\text{ii}} = \frac{2^{7/2} \varepsilon_{\text{S}} (k_{\text{B}} T)^{3/2}}{\pi^{3/2} e^3 (m_{\text{eff}} m_{\text{e}})^{1/2} N_{\text{I}} G(b)} 
    \label{eq:mobility_ionized_impurities}
\end{equation}

with the Boltzmann constant $k_{\text{B}}$, the temperature $T$, the ionized impurity density $N_{\text{I}}$ and the factor $G(b)$ being $G(b) = \text{ln}(b + 1) - [b / (b + a)]$ with

\begin{equation}
    b = \frac{24 \pi m_{\text{eff}} m_{\text{e}} \varepsilon_{\text{S}} (k_{\text{B}} T)^{2}}{e^2 h^2 N_{\text{I}}} 
    \label{eq:mobility_ionized_impurities_b_factor}
\end{equation}

with the Planck constant $h$.

Finally, for the lattice contribution to the mobility, a value of $\mu_{\text{latt}} = 1080 \,  \frac{\text{cm}^{2}}{\text{Vs}}$ is used for n-type c-GaN according to the case of $N_{\text{dsl}} = 0$ and the minimum impurities in \cite{schikora2002molecular}.

\begin{figure}[htb!]
\centering
    \includegraphics[width=0.5\textwidth]
    {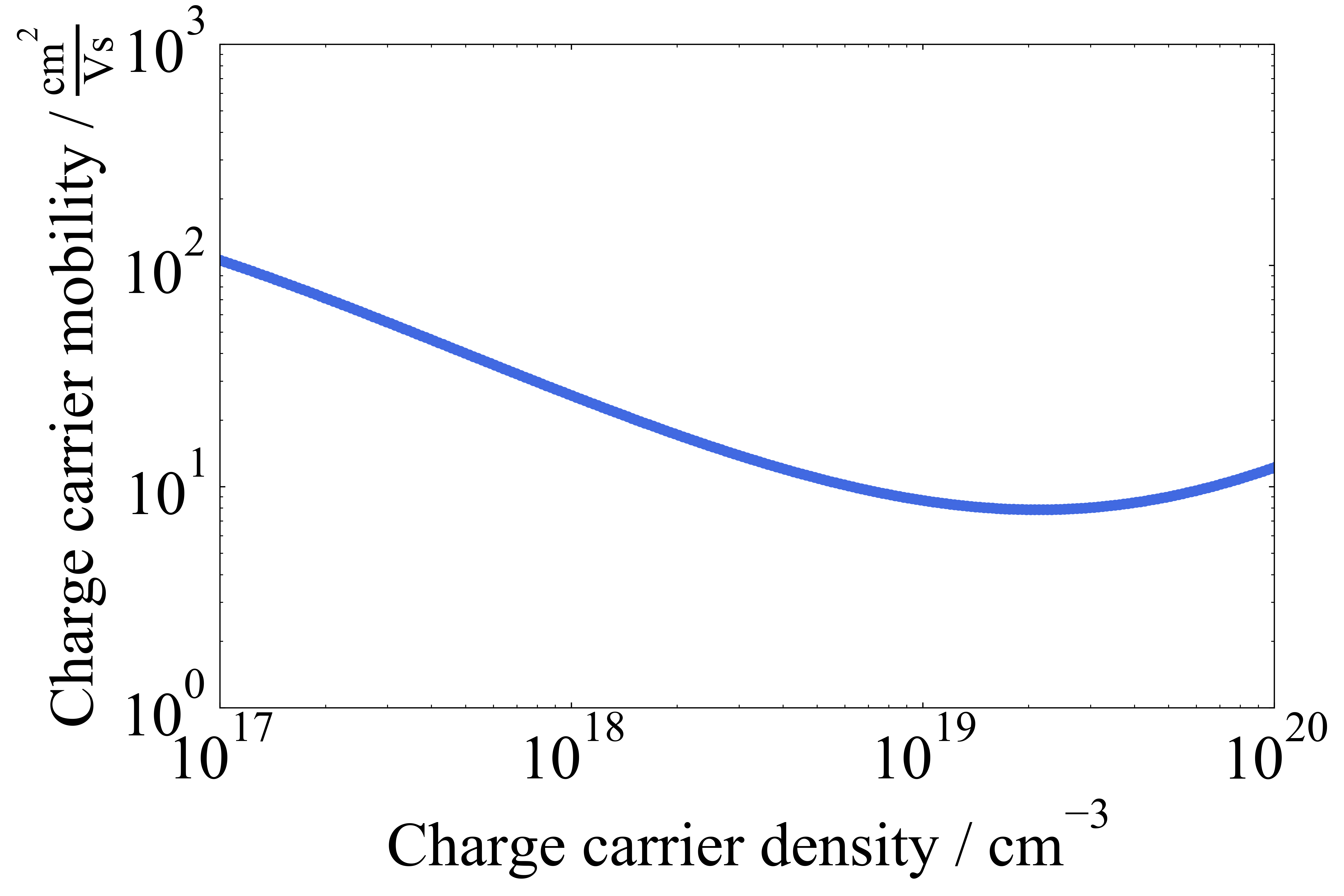}
    \caption{Mobility model for c-GaN used throughout the following evaluations according to Brooks-Herring with N$_{\text{dsl}}$ = 10$^{9}$ \, cm$^{-2}$ assumed in the range of charge carrier densities covered by the samples in this work.}
    \label{fig:mobility_model}
\end{figure}

Resulting from this, the charge carrier mobility of c-GaN is modeled in the range of charge carrier densities covered by the samples listed in Tab.~\ref{tab:sample_information} and shown in Fig.~\ref{fig:mobility_model}. Here, a dislocation density of $N_{\text{dsl}} = 10^{9} \, \text{cm}^{-2}$ is assumed as this value had been determined for comparable samples \cite{zscherp2022aln}. The transition from p-type to n-type conduction of the material is modeled as given in \cite{schikora2002the}. However, due to the low dislocation density compared to \cite{schikora2002the}, we remain in the case of n-type conducting c-GaN for each of our samples. The counterintuitive slight rise in mobility towards high doping levels is due to the model possibly neglecting further scattering types \cite{kosina1998ionized}. Furthermore, the effective mass at the $\Gamma$ point for n-type c-GaN is assumed to be 0.13 \cite{levinshtein2001properties}.

The evaluation of a reflection TDS measurement can be done as described in \cite{hennig2026accessibility} modeling each layer, choosing open parameters for the evaluation with respective optimization boundaries, simulating the terahertz wave-material interaction using Rouard's method \cite{krimi2016non} and finally optimizing the parameters to be determined, which in our case will be the charge carrier densities of the c-GaN and 3C-SiC layers.

\section{Results: Thickness characterization by FTIR}

The thinner the considered layers are, the stronger is the impact of slight changes in layer thickness on the result of the charge carrier density determination \cite{hennig2026accessibility}. Therefore, it is desirable to know this quantity as precisely as possible in order to avoid the need to determine it from TDS measurements as well. Thereby the number of open parameters to be determined is kept as low as possible to allow for a focus of the optimization on the electric quantities at that point. Therefore, Fourier-transform infrared spectroscopy (FTIR) measurements of the samples are executed in advance of the TDS investigation aiming to determine the layer thicknesses of both the c-GaN and the 3C-SiC layers. These are the two most relevant ones given that the AlN layer is negligibly thin and the Si layer is barely probed by the TDS measurement due to the high doping of the layers on top of it. The FTIR measurements are conducted with a Bruker Vertex 70v in combination with a Hyperion 3000 microscope with a Cassegrain configuration and 20x magnification in reflection geometry.

\begin{figure}[htb!]
\centering
    \includegraphics[width=0.98\textwidth]
    {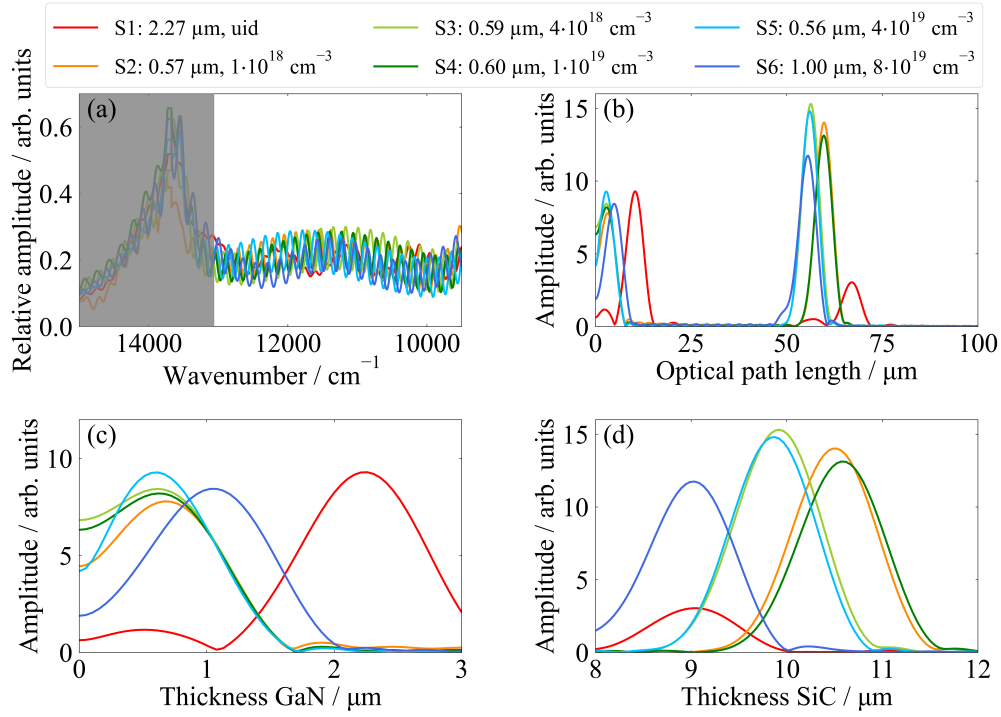}
    \caption{FTIR measurements and evaluation steps of the samples investigated: (a) Complete measurements spanning from 15,000~cm$^{-1}$ to 9,500~cm$^{-1}$, (b) FFT of the measurements indicating the complete optical path lengths, (c) evaluated GaN layer thicknesses, and (d) evaluated 3C-SiC layer thicknesses.}
    \label{fig:FTIR_meas_and_eval}
\end{figure}

The amplitude of the FTIR measurements of the six samples relative to a reference measurement on gold is shown in Fig.~\ref{fig:FTIR_meas_and_eval} (a). Each measurement is saved as the average of 32 scans with a resolution of 8 cm$^{-1}$. While the whole spectrum recorded covers a wavenumber range from 15,000 cm$^{-1}$ to 9,500 cm$^{-1}$, only the range from approx. 13,000 cm$^{-1}$ to 9,500 cm$^{-1}$ is used for further evaluation as it exhibits clear Fabry-Pérot resonances in contrast to the rest of the spectrum marked in gray. Applying a Fast Fourier transform (FFT) to the spectra reveals peaks corresponding to the optical path lengths of the layers, as shown in Fig.~\ref{fig:FTIR_meas_and_eval} (b). The topmost c-GaN layer thickness is retrieved by converting the optical path length into a thickness by dividing the peak position by a factor of 2 to account for two passages due to the reflection geometry and by dividing by the infrared (IR) refractive index of c-GaN $\tilde{n}_{\text{c-GaN IR}} = \sqrt{5.35} \approx 2.31$ \cite{mirjalili1998far}. The resulting peaks corresponding to the GaN layer thickness are shown in Fig.~\ref{fig:FTIR_meas_and_eval} (c). To retrieve the 3C-SiC layer thickness, two times the optical path length of the c-GaN layer is subtracted from the peak positions at longer optical path lengths, before it is divided by a factor $2 \cdot ( \tilde{n}_{\text{3C-SiC IR}} = 2.55)$ \cite{pluchery2012infrared}, analogously to the c-GaN layer. The resulting peaks indicating the SiC layer thicknesses are shown in Fig.~\ref{fig:FTIR_meas_and_eval} (d).

The results of thickness determination using FTIR measurements are summarized and compared to the specifications of the samples in Tab.~\ref{tab:d_results}. The determined GaN layer thickness is realistically close to the original estimation. Only for the weakly doped sample S2, the deviation is comparably high with about 18~\%. For the other samples, the deviation is in the range from 0.5~\% to 6~\% and therefore reasonable. For the SiC layers, the manufacturer of the commercially available substrates states a 10~\% error range around 10.3~\textmu m. Most of the results are well inside this given range. Only samples S1 and S6 are slightly outside of it. 

It is worth noticing that sample pairs S1 and S6, S2 and S4, as well as S3 and S5 each are grown on substrates originating from the same substrate wafer, respectively. This corresponds well to the SiC layer thicknesses of each of these pairs being quite similar, which is a reasonable observation indicating that these results are robust. 

\begin{table}[htbp]
\caption{Comparison of the specified layer thicknesses and the layer thicknesses determined from the FTIR measurements both for GaN and for 3C-SiC layers.}
  \label{tab:d_results}
  \centering
\begin{tabular}{ccccc}
\hline
Number & $d_{\text{GaN Spec}}$ / \textmu m & $d_{\text{GaN FTIR}}$ / \textmu m & $d_{\text{SiC Spec}}$ / \textmu m & $d_{\text{SiC FTIR}}$ / \textmu m \\
\hline
S1 & 2.27 $\pm$ 0.10 & 2.26 & 10.3 $\pm$ 1.0 & 9.04\\
S2 & 0.57 $\pm$ 0.02 & 0.67 & 10.3 $\pm$ 1.0 & 10.52\\
S3 & 0.59 $\pm$ 0.02 & 0.61 & 10.3 $\pm$ 1.0 & 9.93\\
S4 & 0.60 $\pm$ 0.02 & 0.62 & 10.3 $\pm$ 1.0 & 10.57\\
S5 & 0.56 $\pm$ 0.02 & 0.62 & 10.3 $\pm$ 1.0 & 9.87\\
S6 & 1.00 $\pm$ 0.10 & 1.06 & 10.3 $\pm$ 1.0 & 8.98\\
\hline
\end{tabular}
\end{table}

\section{Results: Charge carrier density characterization by terahertz TDS}


Terahertz time-domain spectroscopy measurements of the six c-GaN samples specified in Tab.~\ref{tab:sample_information} are carried out in reflection geometry using a TDS system with mechanical delay line at a rate of 50~Hz according to the setup shown in Fig.~\ref{fig:TDS}. Each sample is measured 100 times after a respective reference measurement on a metal plate is carried out. Then a new reference measurement is recorded followed by an additional 100 sample measurements, each after adjusting the measurement head, making sure the sample is in the focal plane and a possible tilt of the optics is minimized. This procedure is repeated ten times, resulting in 1,000 sample measurements for each sample.  

\begin{figure}[htb!]
\centering
    \includegraphics[width=0.98\textwidth]
    {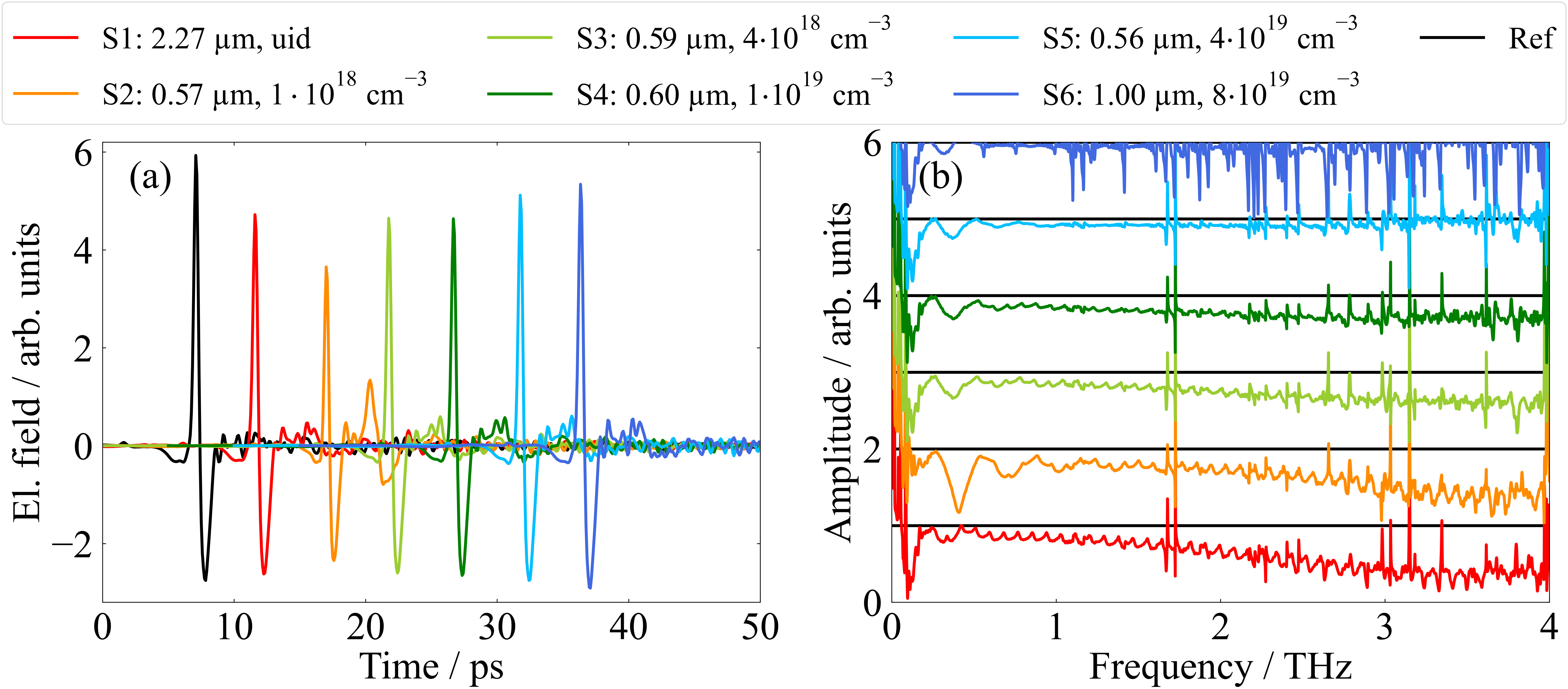}
    \caption{(a) Waveforms of terahertz TDS measurements of each GaN sample investigated. They are shifted in time to allow for a better comparison. (b) Spectral amplitude relative to the reference corresponding to the waveforms. Spectra are shifted along the amplitude axis to allow for a better distinction.}
    \label{fig:measurements_TD_FD}
\end{figure}

Fig.~\ref{fig:measurements_TD_FD} (a) shows a waveform recorded in the time domain for each of the samples as well as a reference and Fig.~\ref{fig:measurements_TD_FD} (b) shows the corresponding spectra relative to the reference spectrum. The spectra are shifted relative to each other. For comparison, each spectrum has its own reference (i.e., unity and its integer multiples due to being relative to itself) plotted in black above it for better comparison. One can see the qualitative trend of a rise in amplitude with an increase in charge carrier density. The samples have a small lateral size of only about 10~mm by 10~mm, of which not even the whole area is covered by the c-GaN layer. Because of this and the terahertz spot size in the focal point being heavily dependent on the wavelength, the low frequencies corresponding to long wavelengths and therefore large focal spot sizes are likely not focused onto the sample completely. Therefore, for the following evaluation, the spectra are high-pass filtered by a Hamming window centered around the frequency $f = 0.5$~THz with a bandwidth of $0.5$~THz. This is done for all samples, except for S2, for which one can clearly see a second peak in the time domain and the corresponding peak due to interference in the frequency domain. The latter peak comes from reflections from the table, which the sample is placed on, because this sample is only a quarter of the size of the others. Therefore, the high-pass filter is shifted to $f = 1.0$~THz for this sample to avoid an influence of these artifacts in the evaluation.

Additionally, the time-domain signal is zero-filled \cite{smith2011terahertz} to a length of 1,000~ps before the FFT corresponding to a high frequency resolution of 1~GHz after the FFT. Then, the spectrum is high-pass filtered as described above and evaluated according to \cite{hennig2024wide, hennig2025simultaneous, hennig2026accessibility} with the charge carrier densities of both GaN and SiC layers as parameters open for optimization with the optimization boundaries in the ranges 

\begin{equation}
    10^{16} \, \text{cm}^{-3} < n_{\text{GaN}} < 10^{22} \, \text{cm}^{-3} \,\,\, \text{and} \,\,\, 10^{17} \, \text{cm}^{-3} < n_{\text{SiC}} < 10^{19} \, \text{cm}^{-3}.
    \label{eq:Optimization_boundaries}
\end{equation}

\begin{table}[htbp]
\caption{Material parameters $\varepsilon_{\text{s}}$,  $m_{\text{eff}}$, and $\mu$ taken from literature for evaluation of the TDS measurements. Additionally, their respective references for the various materials, which the samples are composed of, are given.}
  \label{tab:Material_parameters}
  \centering
\begin{tabular}{c|c|c|c}
\hline
Material &  $\varepsilon_{\text{s}}$ / 1 & $m_{\text{eff}}$ / 1 &  $\mu(N)$ / $\frac{\text{cm}^{2}}{\text{Vs}}$\\
\hline
c-GaN & $9.7$ \cite{levinshtein2001properties} & 0.13 \cite{levinshtein2001properties} & $(\frac{1}{\mu_{\text{dsl}}} + \frac{1}{\mu_{\text{ii}}} + \frac{1}{\mu_{\text{latt}}})^{-1}$, c.f. Eq. 2-5\\ 
AlN & $8.5$ \cite{levinshtein2001properties} & $0.4$ \cite{xu1993electronic} & 300 \cite{chin1994electron}\\
3C-SiC & $9.72$ \cite{choyke1970luminescence} & $0.313$ \cite{yugami1987characterization} & $\frac{330 \cdot (n/3\cdot10^{16})^{0.6} + 650}{1 + (n/3\cdot10^{16})^{0.8})}$ \cite{roschke2001electron}\\
Si & $11.68$ \cite{dai2004terahertz} & $0.26$ \cite{jeon1997nature} & $\frac{1330 - 65}{1 + (n / 8.5 \cdot 10^{16})^{0.72}} + 65$ \cite{caughey1967carrier}\\
\hline
\end{tabular}
\end{table}

For the evaluation according to \cite{hennig2026accessibility}, the static dielectric constant $\varepsilon_{\text{s}}$, the effective mass $m_{\text{eff}}$ and the mobility $\mu$ need to be assumed as material parameters for each material of the samples. For that purpose, the values taken from literature are listed together with their respective references in Tab. \ref{tab:Material_parameters}. 

The results of the evaluation of the 1,000 measurements for each sample are summarized in Tab.~\ref{tab:N_results}. For both $n_{\text{GaN}}$ and $n_{\text{SiC}}$, the mean value of the optimization is given plus/minus the standard deviations of all successful measurements of the respective sample as an error estimation. Measurements for which the optimization reached a limit of the boundaries given in Eq. \ref{eq:Optimization_boundaries}, are not considered successful and therefore neglected. For comparison, the specified Ge concentrations $N_{\text{Ge}}$ are given. For $n_{\text{SiC}}$, the manufacturer named a vague value of few 10$^{17}$ cm$^{-3}$ as the target doping. Compared to this target value, the TDS measurement tends to overestimate the charge carrier density of this 3C-SiC substrate layer. However, the more interesting layer is the c-GaN layer. For these layers, the results of the samples S3 - S6 agree well with the ToF-SIMS estimation with the error estimations overlapping. However, for samples S1 and S2, the results disagree with the specifications. 

\begin{table}[t]
\caption{Comparison of the Ge concentration $N_{\text{Ge}}$ of the c-GaN layers from ToF-SIMS measurements and the charge carrier densities $n_{\text{GaN}}$ obtained from TDS measurements. Additionally the charge carrier densities $n_{\text{SiC}}$ of the 3C-SiC layers are listed as determined simultaneously from the same TDS measurements.}
  \label{tab:N_results}
  \centering
\begin{tabular}{c|c|cc}
 & ToF-SIMS & \multicolumn{2}{c}{TDS} \\  
\hline
Number &  $N_{\text{Ge}}$ / cm$^{-3}$ &  $n_{\text{GaN}}$ / cm$^{-3}$ &  $n_{\text{SiC}}$ / cm$^{-3}$\\
\hline
S1 & uid & 1.5$\cdot$10$^{18}$ $\pm$ 4.3$\cdot$10$^{16}$ & 1.2$\cdot$10$^{18}$ $\pm$ 7.2$\cdot$10$^{16}$\\ 
S2 & 1$\cdot$10$^{18}$ $\pm$ 1$\cdot$10$^{18}$ & 4.9$\cdot$10$^{18}$ $\pm$ 5.3$\cdot$10$^{17}$ & 7.0$\cdot$10$^{17}$ $\pm$ 5.3$\cdot$10$^{16}$\\
S3 & 4$\cdot$10$^{18}$ $\pm$ 2$\cdot$10$^{18}$ & 9.2$\cdot$10$^{18}$ $\pm$ 7.7$\cdot$10$^{18}$ & 7.7$\cdot$10$^{17}$ $\pm$ 5.3$\cdot$10$^{17}$\\
S4 & 1$\cdot$10$^{19}$ $\pm$ 4$\cdot$10$^{18}$ & 9.4$\cdot$10$^{18}$ $\pm$ 6.2$\cdot$10$^{18}$ & 9.7$\cdot$10$^{17}$ $\pm$ 2.7$\cdot$10$^{17}$\\
S5 & 4$\cdot$10$^{19}$ $\pm$ 2$\cdot$10$^{19}$ & 6.2$\cdot$10$^{19}$ $\pm$ 7.1$\cdot$10$^{18}$ & 1.4$\cdot$10$^{18}$ $\pm$ 1.6$\cdot$10$^{18}$\\
S6 & 8$\cdot$10$^{19}$ $\pm$ 2$\cdot$10$^{19}$ & 8.3$\cdot$10$^{19}$ $\pm$ 5.5$\cdot$10$^{18}$ & 4.5$\cdot$10$^{18}$ $\pm$ 3.2$\cdot$10$^{18}$\\
\hline
\end{tabular}
\end{table}

\begin{figure}[ht!]
\centering
    \begin{subfigure}{0.49\textwidth}
        \centering
        \includegraphics[width=\linewidth]{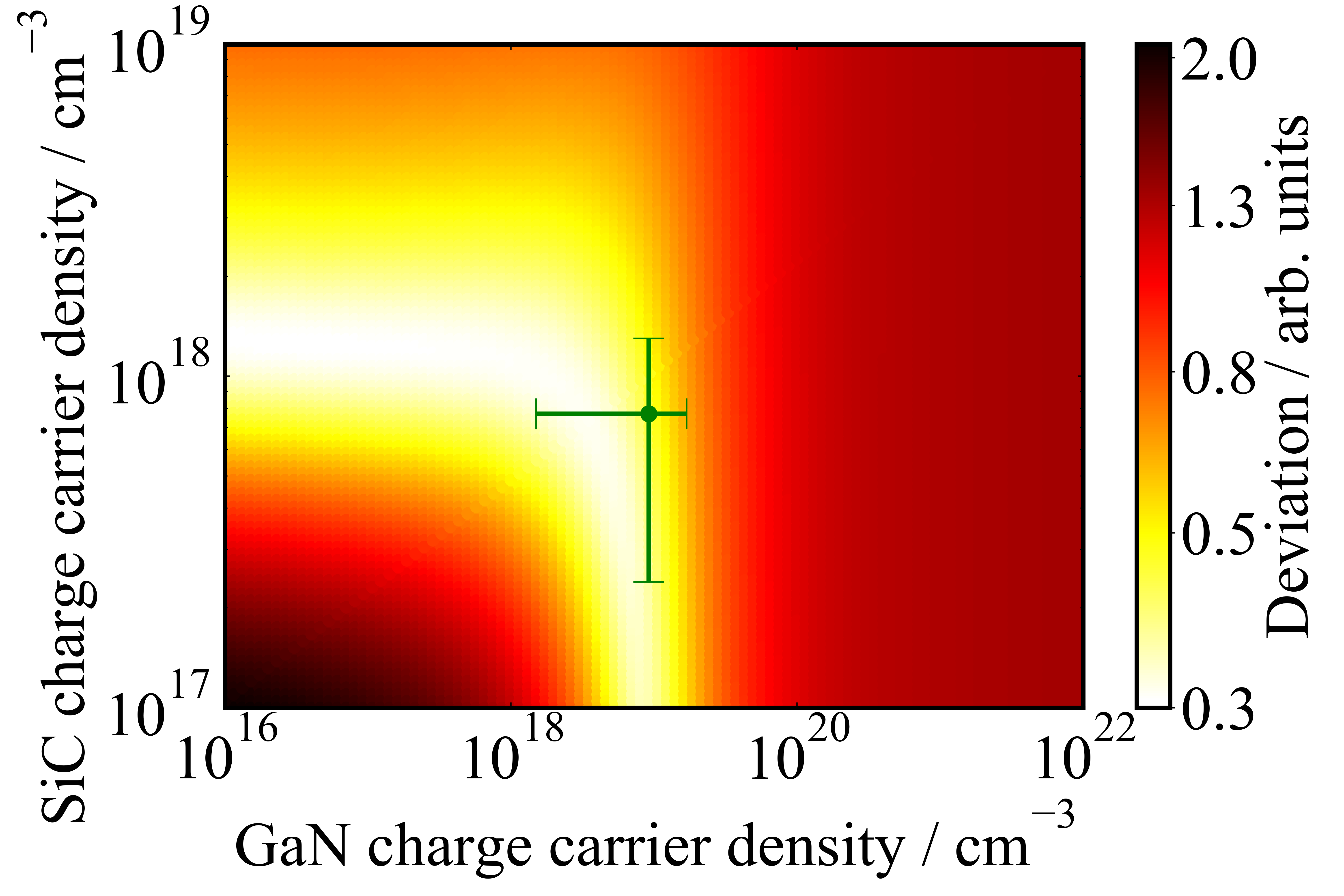}
    \end{subfigure}
    \hfill
    \begin{subfigure}{0.49\textwidth}
        \centering
        \includegraphics[width=\linewidth]{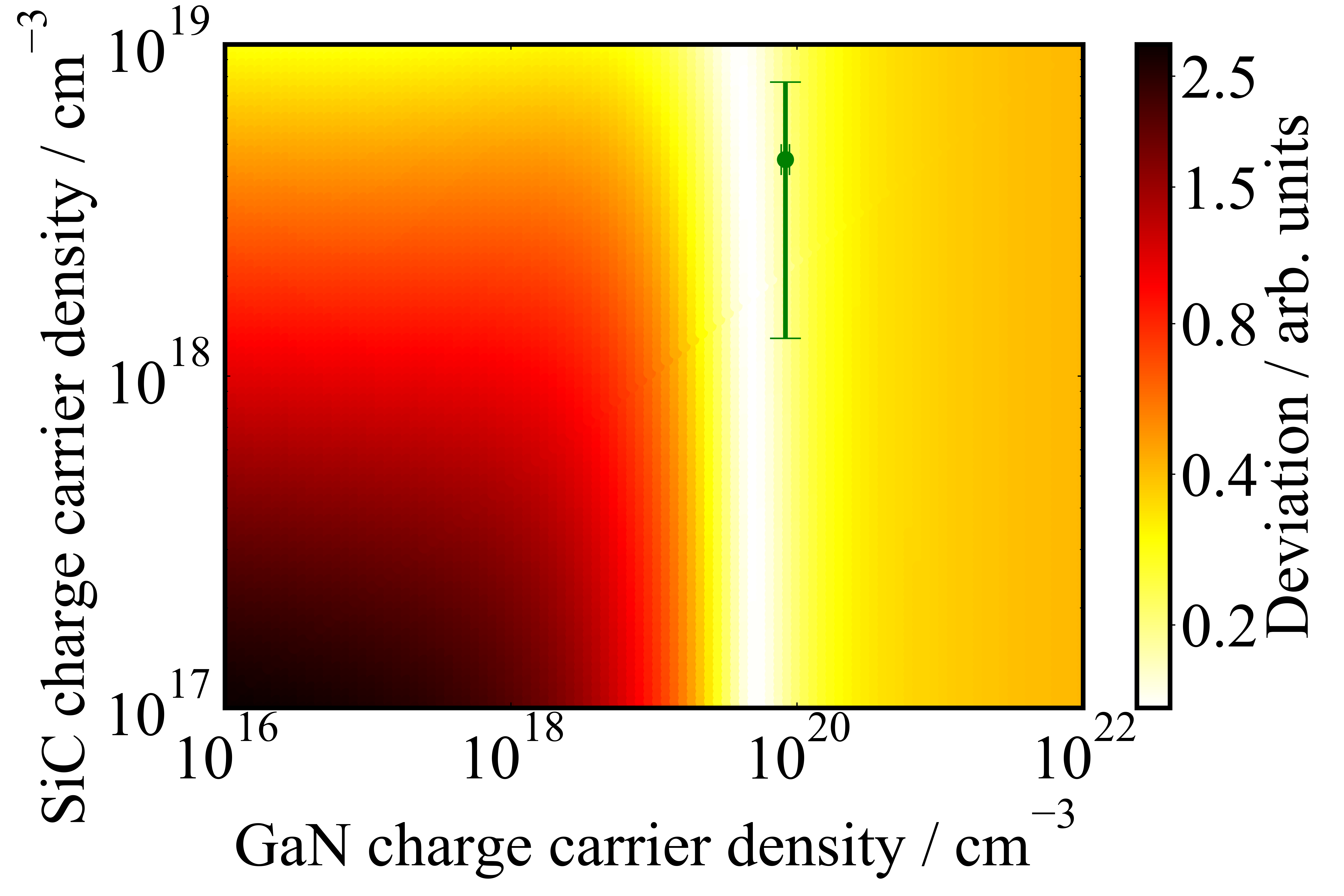}
    \end{subfigure}
    \put(-346,109){(a)}
    \put(-152,109){(b)}
    \put(-244,109){S3}
    \put(-49,109){S6}
    \caption{Heat maps showing error landscapes of the optimization. For simulated spectra based on various assumed combinations of $n_\text{GaN}$ and $n_\text{SiC}$, the deviation from the measured sample spectrum is depicted (a) for the lightly-doped sample S3 and (b) for the highly-doped sample S6. The optimization attempts to find the minimum of the deviation by optimizing these two quantities. The mean and standard deviation of the respective evaluation results are plotted for each sample inside the heat map in green.}
    \label{fig:Heatmaps_Optimization}
\end{figure}

In order to better understand these results, the error landscape of the optimization is considered. Figs.~\ref{fig:Heatmaps_Optimization} (a) and (b) show the error landscapes for the samples S3 and S6, respectively. The heat maps show the deviations between a measured and various simulated spectra on the color bar. For these heat maps, terahertz spectra after the interaction with samples of 10,000 different combinations of the two optimized quantities $n_{\text{GaN}}$ and $n_{\text{SiC}}$ are simulated, covering the whole ranges given in Eq. \ref{eq:Optimization_boundaries}. The deviation is computed by the same criterion as in the optimization

\begin{equation}
    \text{Deviation} = \sqrt{\frac{\sum_{i=1}^{l}(|a_i|-|b_i|)^2}{l}},
    \label{eq:Error_landscape_deviation}
\end{equation}

which is a root-mean square criterion, comparing the absolute value of the two spectra $|a_i|$ and $|b_i|$ at each frequency position $i$. 

\begin{figure}[bp!]  
\centering
    \includegraphics[width=0.85\textwidth]
    {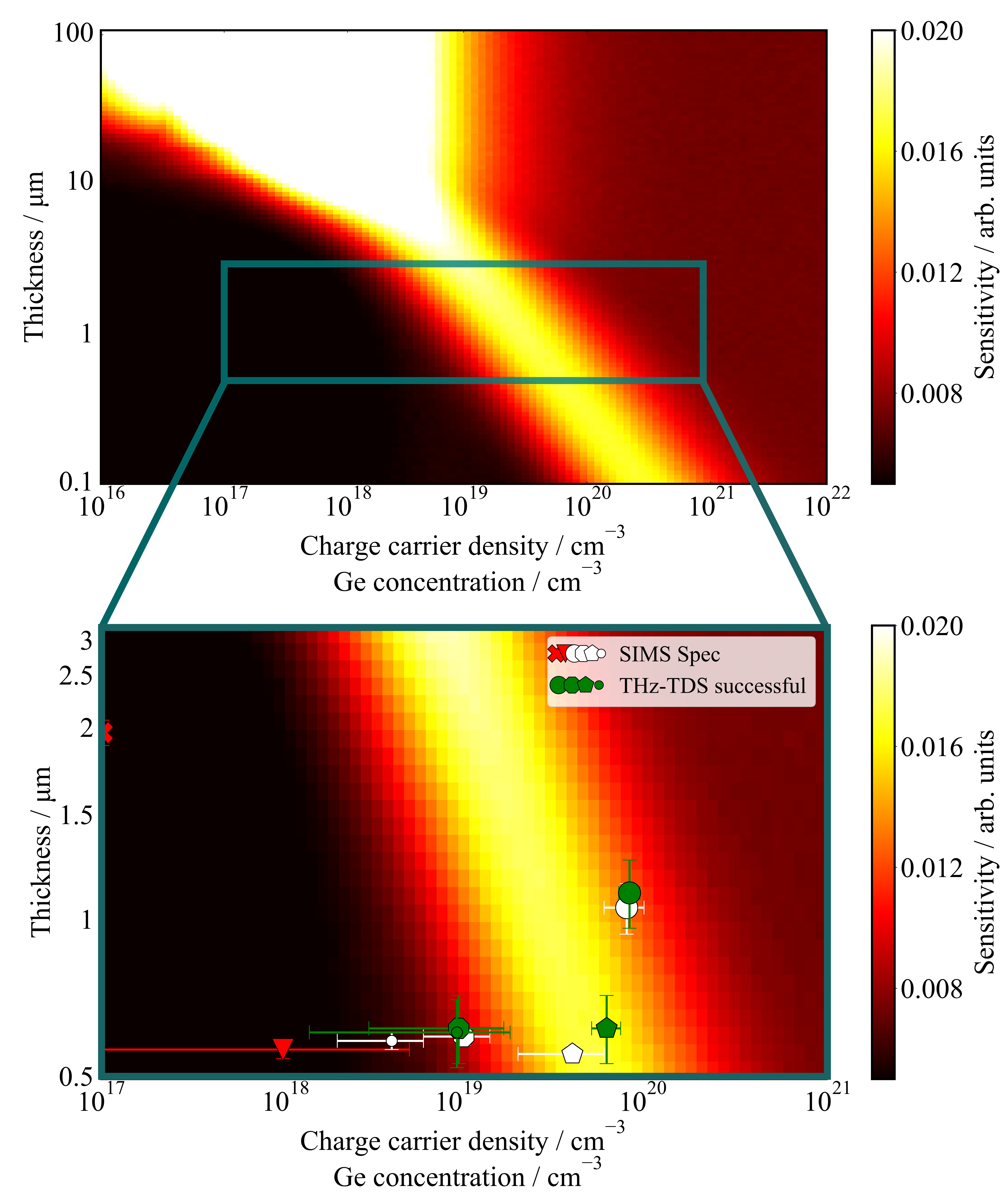} 
    \put(-288,365){(a)} 
    \put(-288,173){\textcolor{white}{(b)}}
    \caption{(a) Sensitivity heat map for cubic GaN according to \cite{hennig2026accessibility} and (b) a zoom into the region of this heat map relevant for the c-GaN samples investigated. Additionally, the reference values of the thickness and Ge-concentration as well as the results for the thicknesses and charge carrier densities of the samples are sketched inside the heat map. The error bars indicate an assumed range of 10~\% as an error interval for the thicknesses determined by FTIR and show the standard deviation for the charge carrier density determined by TDS measurements.}
    \label{fig:Heatmap_results_TDS}
\end{figure}

On the one hand, Fig.~\ref{fig:Heatmaps_Optimization} (a) shows that for the lightly-doped sample S3 a fairly large area of combinations of $n_{\text{GaN}}$ and $n_{\text{SiC}}$ leads to a minimal deviation between the measurement and the simulations, indicating that they both, and especially $n_{\text{GaN}}$ could have a value spanning over multiple orders of magnitudes. This area marked in white in the heat map is the area which the optimization will find a minimum in, resulting in a relatively wide-spread range of values for $n_{\text{GaN}}$ for S3, corresponding to the large standard deviation in Tab.~\ref{tab:N_results}. This can be explained in the following way: This GaN layer is optically so thin that for the large-wavelength terahertz radiation it is almost perfectly transparent, barely influencing the measurement. However, the thicker doped 3C-SiC layer is probed, which is in good agreement with the fact that 4H-SiC with comparable optical parameters in that doping range and a thickness of around 10~\textmu m can be characterized by terahertz TDS \cite{hennig2025simultaneous}.

On the other hand, Fig.~\ref{fig:Heatmaps_Optimization} (b) shows that for the most highly-doped sample S6, the deviation between the measurement and a simulation is minimal for a well defined value of $n_{\text{GaN}}$, while $n_{\text{SiC}}$ could vary over a wide range of values without having a noticeable impact on the deviation between measurement and simulation. This explains the comparably small standard deviation for $n_{\text{GaN}}$ and the comparably large standard deviation for $n_{\text{SiC}}$ for S6 in Tab.~\ref{tab:N_results}. This behavior can be explained in the way that the c-GaN layer of this sample is optically thick enough to have a noticeable impact on the reflected terahertz pulse. However, it is optically too thick to allow for a characterization of the 3C-SiC layer below with the remaining transmitted remainder of the terahertz pulse. These two error landscapes in Fig. \ref{fig:Heatmaps_Optimization} represent the two extreme examples reached by the set of samples investigated, while the other successfully characterized samples are in between these two cases. The samples S1 and S2 not characterized successfully suffer even more from the vanishing impact of their even more lightly-doped GaN layers on the TDS measurement, making the measurement sensitive only to the substrate.



The fact that a highly-doped semiconductor layer on a doped substrate can be characterized by the presented approach of reflection terahertz TDS and an amplitude-based evaluation of the spectra, even if it is very thin, was predicted by simulations before \cite{hennig2026accessibility}. Here, we show that the charge carrier density even of a sub-\textmu m thin sample can be determined this way. In order to contextualize the investigated c-GaN samples in such simulations, a sensitivity heat map for the present four-layer sample setup is simulated for c-GaN layer thicknesses between 0.1~\textmu m~<~$d_{\text{GaN}}$~<~100~\textmu m and charge carrier densities $10^{16} \, \text{cm}^{-3}~<~n_{\text{GaN}}~<~10^{22} \, \text{cm}^{-3}$ according to \cite{hennig2026accessibility} and plotted in Fig. \ref{fig:Heatmap_results_TDS} (a). Fig. \ref{fig:Heatmap_results_TDS} (b) shows a zoom into the relevant area of this simulation including the specifications as well as the results for $n_{\text{GaN}}$ from the successful TDS evaluations and $d_{\text{GaN}}$ from the FTIR measurements. In this heat map, it becomes obvious that samples S1 and S2 are outside the range, terahertz TDS is sensitive in. The other samples are within this range and, therefore, $n_{\text{GaN}}$ can be determined for them successfully.




\section{Conclusion and outlook}

With these results, the ability of terahertz time-domain spectroscopy as a technique to determine the charge carrier density as one of the most relevant properties of Ge-doped c-GaN layers is proven experimentally in a range covering two orders of magnitude roughly from $1~\cdot~10^{18}$~cm$^{-3}$ to $1~\cdot~10^{20}$~cm$^{-3}$, even for sub-\textmu m-thin samples. These results are verified by ToF-SIMS measurements of the Ge concentration. Additionally, the limits of terahertz TDS in terms of layer thickness and doping density have been characterized by means of simulations for this material. Therefore, terahertz TDS in combination with FTIR for thickness characterization are powerful techniques to determine the charge carrier density of sub-\textmu m-thin c-GaN layers. In the future, these techniques could allow for a fast and completely contactless and non-destructive characterization of the samples with measurement rates up to the kilohertz range, right after or even during the growth process. It underlines the possibility to integrate terahertz TDS as a measurement technique in the production process of c-GaN in possible optoelectronic applications in the future with the potential of both rapid and exhaustive characterization of all samples, eliminating costly and time-consuming contacting of exemplarily chosen samples.

\section{Back matter}
\begin{backmatter}
\bmsection{Funding}
This project is supported by the Federal Ministry for Economic Affairs and Climate Action (BMWK) on the basis of a decision by the German Bundestag.

\bmsection{Acknowledgment}
-

\bmsection{Disclosures}
The authors declare no conflicts of interest.

\bmsection{Data Availability Statement}
Data underlying the results presented in this paper are not publicly available at this time but may be obtained from the authors upon reasonable request.


\end{backmatter}

\bibliography{bib}

\end{document}